## ARTICLE

Check for updates

# Magnetically powered metachronal waves induce locomotion in self-assemblies

Ylona Collard [1 ✉], Galien Grosjean [1,2] & Nicolas Vandewalle[1]

When tiny soft ferromagnetic particles are placed along a liquid interface and exposed to a vertical magnetic field, the balance between capillary attraction and magnetic repulsion leads to self-organization into well-defined patterns. Here, we demonstrate experimentally that precessing magnetic fields induce metachronal waves on the periphery of these assemblies, similar to the ones observed in ciliates and some arthropods. The outermost layer of particles behaves like an array of cilia or legs whose sequential movement causes a net and controllable locomotion. This bioinspired many-particle swimming strategy is effective even at low Reynolds number, using only spatially uniform fields to generate the waves.

[1] GRASP, Department of Physics, University of Liège, Allée du 6 Août 19, B4000 Liège, Belgium. [2] Waitukaitis Group, IST Austria, Am Campus 1, 3400 Klosterneuburg, Austria. ✉email: ycollard@uliege.be

In recent years, many efforts have been made to create synthetic microswimmers. These tiny structures use a wide variety of techniques to move in a fluid at low Reynolds number. Some mimic living microorganisms with artificial flagella[1,2] or rotating helices[3–5] while others are propelled by chemical reactions[6,7], ultrasound[8,9], or even light[10,11]. The motivations for studying these systems range from the fundamental understanding of biological processes to the development of medical and technological applications[12,13]. However, the synthesis, manipulation, and assembly of the required microscopic components can be a challenge. One possible approach to facilitate the fabrication process is to rely on self-assembly[14–16].

For instance, floating crystals of particles can form owing to a magnetic repulsion between particles in a confinement[17,18] or through a combination of a magnetic repulsion and an attraction due to capillary forces[19,20]. The latter, magnetocapillary self-assemblies, are the focus of this paper. Similar-looking structures can also be obtained through dynamic self-assembly under a constant supply of energy, for example, with a combination of magnetic attractions and hydrodynamic repulsions between rapidly rotating disks[21], or through a combination of magnetic, hydrodynamic, and capillary forces with magnetic droplets under precessing fields[22]. When exposed to a spatially uniform, time-varying magnetic field, magnetocapillary self-assemblies have been shown to spontaneously move along the interface[15]. This intriguing behavior has since been the subject of numerous studies[23–29]. No general approach for swimming has yet been developed, though, as this phenomenon is highly dependent on the geometry of the particular assembly[24,26].

To overcome this problem, we can find inspiration in nature. Microbial swimmers often use one or several dedicated hair-like appendages for propulsion[12]. When these slender organelles operate in large groups, they are called cilia. Motile cilia are found on various eukaryotic cells, including epithelial cells, where they are used to transport fluid or objects inside the body, and on ciliates, a group of unicellular organisms that use them, among other things, for propulsion. The most striking feature of motile cilia is their asynchronous beating. Cilia operate in succession, producing a wave-like motion called a metachronal rhythm. Metachronal waves have been shown to drastically increase swimming speed and efficiency[30,31]. They are also found in the gait or stroke of various invertebrates, including crustaceans[32], worms[33], and insects[34].

To create a net flow and swim, a ciliate must periodically deform in a way that is not time-reversible. This is a necessary condition due to the properties of low Reynolds number flows[12]. This nonreciprocal motion can happen in two ways. First, the individual stroke of a single cilium can obey this condition. And second, the motion of several cilia relative to each other can also break time-reversal symmetry. Indeed, two nearby bodies performing a reciprocal motion can still generate a net displacement if there is a phase difference between them[35].

A ciliary beat is divided in two parts: an effective stroke and a recovery stroke. During the effective stroke, the tip of the cilium is further away from the surface of the ciliate than during the recovery stroke. This means that more fluid is displaced during the effective stroke, which causes a net flow over a complete period. The phase shift between adjacent cilia can induce one or more metachronal waves on the surface of the living organism[36]. These waves emerge from hydrodynamic coupling between the cilia, and possibly as well from coupling through the cell membrane[37]. Therefore, the beating pattern of a cilium has to accommodate both individual and collective effects. Swimming speed, efficiency, and direction are all governed by both the individual strokes and the waves[31].

Various attempts have been made to mimic cilia and their metachronal rhythm. On the one hand, simple artificial cilia can be fabricated on a substrate and actuated, although synchronously, by magnetic fields[38–40]. On the other hand, the macroscopic approach of using individual actuators to generate the metachronal rhythm is not well-suited for miniaturization[41]. Metachronal swimmers that use active materials to produce a traveling wave have also been theorized[42] and successfully implemented, using structured light to locally trigger the wave[1].

The particular structures shown here are formed by self-assembly, through a combination of capillary attraction and magnetic dipole–dipole interactions; however, the same general approach could be applied to microfabricated magnetic cilia. When particles are placed at a water–air interface, they locally deform the interface. This curvature is due to gravity and surface tension, and depends on the shape, buoyancy, and wetting properties of the particles[43]. Nearby particles will therefore attract or repel, as each particle experiences an inclination of the interface caused by the presence of another. Two identical spheres will deform the interface in the same way, inducing an attraction between them. This phenomenon of agglomeration is playfully called the Cheerios effect[44]. In order to control the interaction between the particles, we use soft ferromagnetic spheres[19]. Magnetic dipoles of controllable magnitude and direction can be reversibly induced in the particles using external magnetic fields. To avoid contact between the particles, a vertical magnetic field $\mathbf{B_z}$ is applied perpendicularly to the interface. This magnetic field induces magnetic dipoles in the particles, leading to repulsive interactions. The balance between capillary attraction and magnetic repulsion can lead to an equilibrium distance larger than the particle diameter for typical field values of a few milliteslas[19].

The magnetocapillary interaction between two soft ferromagnetic particles has been investigated in depth in earlier works[23,25]. The distance $r_{ij}$ between two particles $i$ and $j$ can be modified by adjusting the magnetic field $B_z$ as long as contact is avoided. The dimensionless interaction energy between these particles can be described by

$$u_{ij} = -K_0\left(x_{ij}\right) + \frac{\mathrm{Mc}}{x_{ij}^3}, \tag{1}$$

where $K_0$ is a modified Bessel function of the second kind, $x_{ij} = r_{ij}/\lambda$ is the normalized distance in the horizontal plane between beads, using $\lambda = \sqrt{\gamma/\rho g}$ as the capillary length associated with the characteristic interface deformations[44]. The parameter Mc in Eq. (1) is the magnetocapillary number capturing the competition between magnetic and capillary effects. This number is roughly independent of particle sizes[20], such that the results obtained in this paper could be expanded to systems of different scales. Particles ranging from 3 μm to ~1 mm could in theory be bound by the magnetocapillary interaction[25]. This range is determined by gravity, as larger particles would sink and smaller ones would not deform the surface enough to cause a significant capillary attraction[25,43]. The magnetocapillary interaction can lead to the formation of well-ordered floating rafts at a liquid–air interface[19], resulting in a wide variety of self-assembled structures[15,20].

Here, we propose an approach where a spatially uniform, precessing magnetic field triggers waves of deformations on the periphery of these self-assemblies. Depending on the configuration of the field, this can lead to rotational or translational motion. We identify and discuss two types of nonreciprocal deformations caused by these waves, which present similarities with the swimming strategies of ciliates. This wave-based collective approach to locomotion allows to move many-particle rafts at low Reynolds number, and is robust for various assembly configurations.

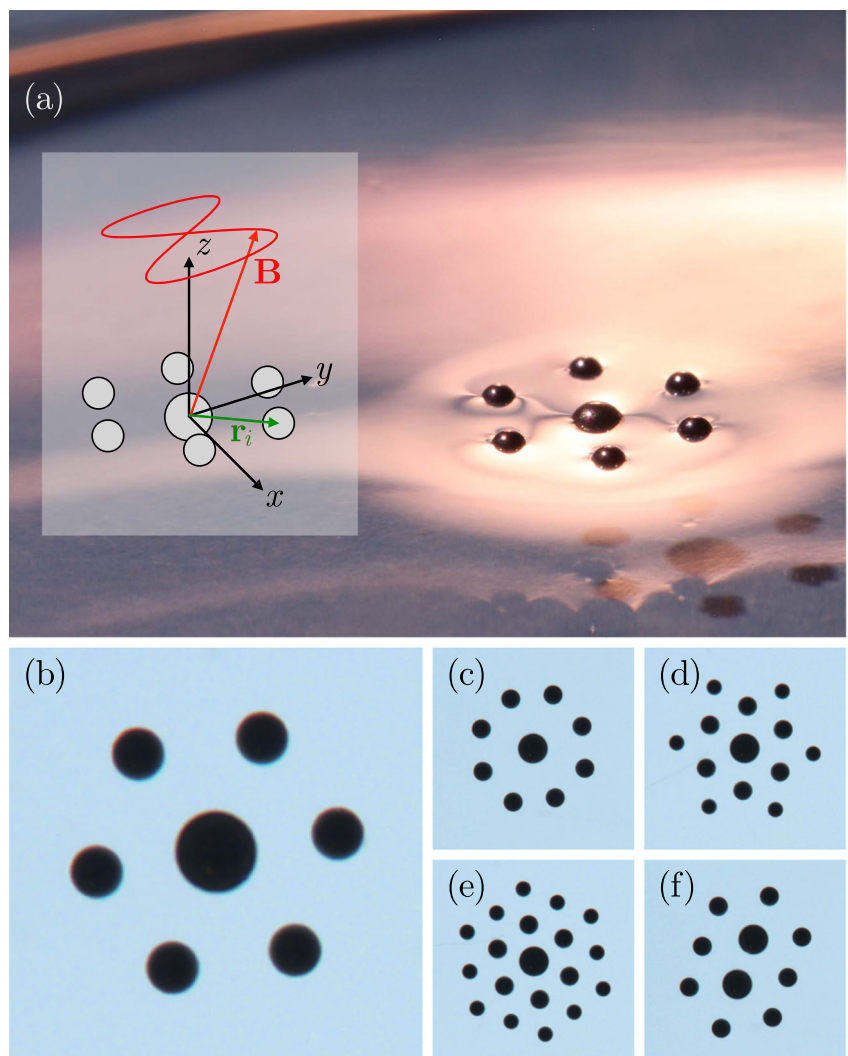

**Fig. 1 Examples of self-assembled rafts.** Typical magnetocapillary assemblies with between 7 and 19 beads. The particle diameters are 400, 500 and 800 μm. **a** A $N = 7$ assembly, with one large and six intermediate particles, or {1,6,0}. The deformation of the liquid around each particle can be seen. The inset shows the frame of reference we adopt for rafts with rotational symmetry, the direction of the magnetic field **B**, and the distance between the central bead and a peripheral one **r**$_i$. The eight-shaped line shows the trajectory of the magnetic field when the vertical component is constant and a time-dependent horizontal component is applied. The self-assemblies shown in this figure are not subject to horizontal magnetic fields but only to a constant vertical magnetic field. **b** Top view of the hexagonal raft {1,6,0}; **c** top view of the octagonal raft {1,8,0}; **d** top view of {1,6,6} composed of particles of three different sizes; **e** top view of {1,6,12}; and **f** top view of the asymmetric raft {2,8,0}.

## Results

### Self-assembly

We focus on self-assembled rafts constituted by $N$ soft ferromagnetic beads of different diameters placed at a water–air interface and immersed in a vertical magnetic field **B**$_z$ perpendicular to the interface. Figure 1 presents a few structures from regular to more complex assemblies. The total number of particles $N$ ranges between 7 and 19, and the diameters $D_1$, $D_2$, and $D_3$ used herein are 800, 500, and 400 μm. To simplify the notations, we will designate by $D$ the diameter of the peripheral beads of an assembly in the following sections. We introduce the notation {$N_1$, $N_2$, $N_3$}, where the $N_i$ are the number of large, intermediate, and small particles, respectively. For instance, Fig. 1a is a {1,6,0} assembly. Figure 1c–f shows several other assemblies we will discuss in this paper. While the equilibrium distance between particles is independent of their size, larger particles are naturally located toward the center of self-assemblies due to gravity[20]. The combination of long-range capillary pull and short-range magnetic repulsion typically leads to compact hexagonal structures, such as in Fig. 1b, d–f. However, the weight of the assembly can curve the water surface downward, which explains why fivefold symmetries are sometimes observed[20]. With a larger particle in the center, sevenfold, and even eightfold symmetries can become stable owing to its stronger capillary pull, as shown in Fig. 1c.

### Mimicking ciliated locomotion

The main objective of this work is to develop a universal strategy to move self-assembled rafts composed of many particles, drawing inspiration from the metachronal rhythms used by ciliates and some arthropods. A common example of ciliated organism is the algae colony *Volvox*, which has a roughly spherical shape. Each individual cilium on *Volvox* follows a periodic beat with an effective and a recovery stroke, as shown in Fig. 2a. The blue cycle shows the position of the tip of the cilium over time. We exploit the fact that peripheral beads in magnetocapillary assemblies can describe a similar flow pattern (Fig. 2b). Ciliary beats are generally quite complex, asymmetric and can either be planar or nonplanar. However, the motion of a rigid sphere near a surface can produce a similar flow pattern, at least in the far field, and is therefore often used to model a cilium[39,45–48]. In the case of an elliptical trajectory, one can show that the flow far from the sphere is proportional to the area of the ellipse projected on the plane perpendicular to the surface[45].

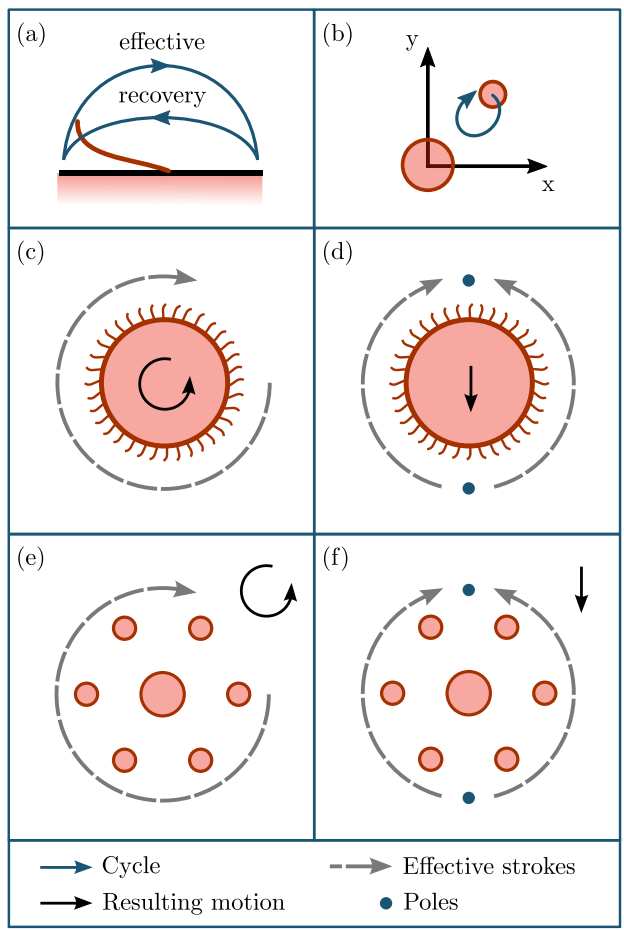

**Fig. 2 Swimming strategy of ciliates compared to magnetocapillary rafts. a** Effective and recovery stroke of a cilium (red), where the blue cycle shows the position of the tip over time. **b** The trajectory of a peripheral bead in a magnetocapillary assembly can also describe a cycle on the horizontal plane. **c** Sketch of *Volvox*. When the effective strokes of the cilia (gray dashed arrow) have the same orientation around the body, *Volvox* rotates (black arrow). **d** When the ciliary beats point toward the same direction, *Volvox* swims straight-ahead (black arrow). Metachronal waves travel from one pole to the other (blue dots) in the same direction. **e** Sketch of an assembly around which a deformation wave travels, causing a rotation. **f** Conversely, two deformation waves propagating around the body cause a translation. The black arrows represent the trajectories of the assembly, the gray dashed arrows represent the directions of the effective strokes and the blue dots represent the poles, where the waves start and end in the translation case.

Remarkably, *Volvox* is able to change its swimming direction by changing the symmetry of the strokes on its body[49]. When it swims in a straight line, metachronal waves propagate between two poles on its body, in the same direction as the effective strokes of the individual cilia[47]. To change direction, for instance when exposed to light, *Volvox* can revert the strokes on part of its body[49]. Figure 2c, d illustrates symmetries on *Volvox* that would lead to pure rotation and pure translation, respectively. The direction of the effective strokes is represented by the gray arrows, while the resulting motions are represented by the black arrow.

This is a greatly simplified view, as the configurations adopted by *Volvox* lie on a continuum. Nonetheless, changing the symmetry of how the strokes are distributed on the body produces different behaviors. By using different symmetries for the variations of the magnetic field, we report in this paper how to change the direction of both the metachronal waves and the strokes, causing different behaviors. This is illustrated in Fig. 2e, f.

Focusing first on the {1,6,0} assembly from Fig. 1b, the central bead is surrounded by a shell of six neighboring beads forming a hexagon. These peripheral particles will play the role of cilia, as sketched in Fig. 2b. To generate metachronal locomotion, these simple cilia must become motile. The energy required to set the cilia in motion, which in eukaryotes is provided by the hydrolysis of adenosine triphosphate, will here come from the dipole–dipole interaction between the beads. Indeed, the addition of a horizontal field, for example, along the *x*-axis, can change the interaction energy. This causes a distortion in the assembly, as it rearranges to find a new minimum of energy. The interaction potential between neighboring beads is now given by

$$u_{ij} = -K_0(x_{ij}) + \frac{Mc(1+\beta^2)}{x_{ij}^3}\left[1 - 3\cos^2\theta_{ij}\right], \qquad (2)$$

where $\beta = B_x/B_z$, and $\theta_{ij}$ is the angle formed by $r_{ij}$ and the external field. When the horizontal component $B_x$ vanishes, $\beta$ goes to zero and $\theta_{ij}$ reaches $\pi/2$, such that Eq. (2) simplifies into Eq. (1). Additional information about this interaction can be found in Supplementary Note 1 and Supplementary Fig. 1.

In Fig. 3, quasi-static experiments have been performed by increasing the horizontal field $B_x$ step by step, in order to observe the distortion of the structure. By tracking the distances $r_i$ between the six beads and the central one (see Fig. 1a), one can build the dimensionless shape distortion vector

$$\sigma = \frac{1}{6\langle r\rangle}\sum_{i=1}^{6} r_i, \qquad (3)$$

which is zero for a regular hexagon. Figure 3a shows $\sigma = |\sigma|$ as a function of the field ratio $\beta$ meaning that the shape deviates from a regular polygon. At a critical ratio $\beta_c \approx 0.17$, the system collapses, i.e., some beads come into contact. Figure 3b presents four snapshots of the assembly for $\beta = 0.0285$, $\beta = 0.0857$, $\beta = 0.1428$, and the collapse at $\beta > \beta_c$. An asymmetric deformation appears along the *x*-axis, causing $\sigma$ to become nonzero as the particles in the direction of $B_x$ ($r_i$ and $B_x$ parallel) move closer to the central one. This causes $\sigma$ to orient in the direction opposite to $B_x$. A slight azimutal reorientation of the assembly is also observed. One could wonder why the interaction is not mirror-symmetric with respect to the *y*-axis, as it is a function of the magnetic field squared. The origin of this asymmetry lies in the vertical positions of the particles. The center of the larger central particle sits deeper in the liquid, so that the angle $\theta$ in Eq. (2) depends on the sense of the vector $B_x$. The continuous curve in Fig. 3a is a simplified model developed from Eq. (2), taking into account the fact that the central bead is larger than the others. More details on the model and the discussion on asymmetry can be found in Supplementary Note 2 and Supplementary Fig. 2.

**Rotational motion**. Because of this asymmetry, the distortion of the assembly follows the direction of the magnetic field. By tracking it, we can follow the metachronal waves around the periphery of the assembly. As with ciliates, both the individual strokes and the phase shift between neighboring particles can produce a net flow. As will be shown later on, both effects would lead to a motion in the direction of the metachronal wave. We can therefore expect different behaviors as a function of the symmetry of the waves, as sketched in Fig. 2. In order to propagate the distortion around the assembly, we will first consider a counterclockwise rotating field along the horizontal plane, while keeping a constant field in the vertical direction. This can be written

$$
\begin{aligned}
B_x &= B_h \cos(2\pi f t), \\
B_y &= B_h \sin(2\pi f t), \\
B_z &= B_v,
\end{aligned}
\tag{4}
$$

with a typical rotation frequency $f$ between $10^{-1}$ Hz and 2 Hz, leading to periodic distortions with period $T = 1/f$. An example trajectory is shown in Fig. 4a, as well as Supplementary Movie 1. Frequencies up to 10 Hz have been tested, as discussed in Supplementary Note 3 and Supplementary Fig. 3. One expects that dipole–dipole interactions along the horizontal plane will induce a distortion propagating on the periphery of the assembly[29]. This distortion is shown in Fig. 4b, where a color scale from yellow to red is used to characterize the distance $r_i$ between each peripheral bead $i$ and the central one. The particle closest to the central one is colored yellow, and the farthest red. By following the yellow bead, one can see that the assembly is rotating along the periphery of the assembly, as denoted by the arrow. In a frame of reference rotating with the assembly, the trajectory of each peripheral bead describes a cycle as sketched in Fig. 2b.

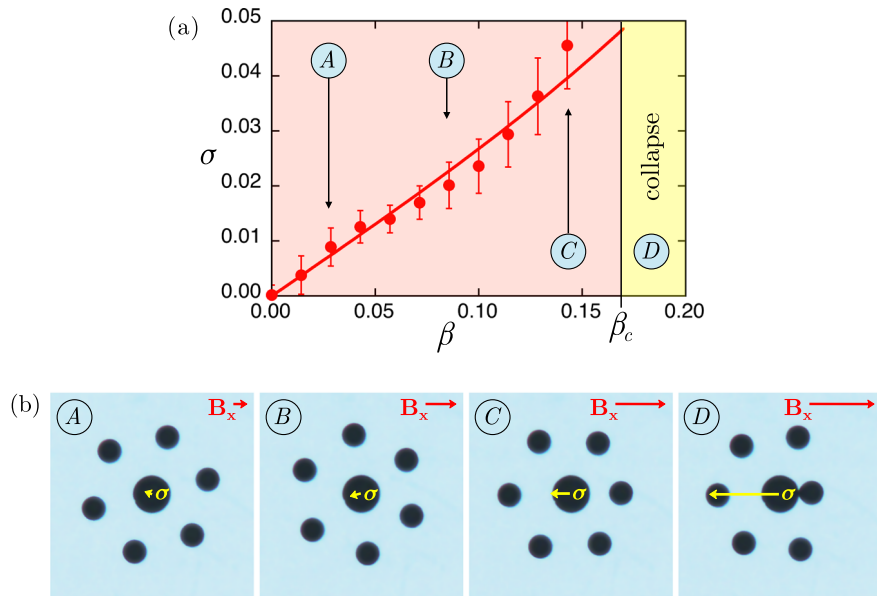

**Fig. 3 Distortion of a hexagonal self-assembly. a** Distortion $\sigma$ increases with the ratio between horizontal ($B_x$) and vertical ($B_z$) magnetic fields $\beta = B_x/B_z$ until the collapse, i.e., the contact between two or more beads at the critical ratio $\beta_c$. Data points correspond to the average over four experiments and the error bars correspond to the standard deviation. The continuous curve is a fit from theory, obtained by balancing magnetic and capillary forces. **b** Visualization of the distortion for {1,6,0} at $\beta = 0.0285$ (A), $\beta = 0.0857$ (B), $\beta = 0.1428$ (C), and $\beta > \beta_c$ (D). The distortion vector $\boldsymbol{\sigma}$ tends to orient in the direction opposite to **$B_x$**.

Figure 4c shows the dimensionless distances $(r_i(t) - \langle r \rangle)/D$, where $\langle r \rangle$ is the distance in the absence of distortion. Each of the six peripheral beads is shown over two periods of the precessing magnetic field. To improve readability, the measurements for each bead were filtered at the frequency corresponding to the maximum of their Fourier transform (see "Methods" section). Each particle oscillates at the frequency of the magnetic field. The minimum of each $r_i(t)$ over one period is denoted by vertical dot dashed lines, emphasizing that the oscillations of the beads are successively shifted by $T/6$. The small variations in amplitude are induced by the small differences in magnetic properties and wetting of the beads. This propagating distortion is the signature of a metachronal wave rotating around the body, in the direction opposite to the effective strokes as sketched in Fig. 2e. This is similar to the symmetry responsible for the rotational motion of *Volvox*, except in that case the wave is in the direction of the working stroke (Fig. 2c).

The resulting motion of the structure is a rotation at an angular speed $< 2\pi f$. This is illustrated in Fig. 4a, where the trajectory of the beads is shown over four periods $T$. This rotational motion comes from the cooperative motion of all beads. Each bead in the perimeter of the structure is roughly displaced by $D/2$ over a period $T$. Note that depending on the frequency and amplitude, different regimes can be observed under a rotating field, as explorer for three particles in ref. [29]. For instance, very low frequencies can lead to a locking to the external field, and for higher frequencies, inertia can start to play a role. A sweep in frequency can be found in Supplementary Fig. 3. Here, we limit our study to the regime where locomotion is due to nonreciprocal deformations.

**Translational motion**. While using a precessing field to generate a rotation of the body seems natural, the symmetry needed to generate a translation is not trivial. In the typical ciliate depicted in Fig. 2d, metachronal waves appear and disappear at two poles on the body. A 2D equivalent would be two waves starting from the back and traveling on each side along the periphery of the assembly to the front of the system. The horizontal field should cause two waves of deformation propagating in the same direction, and vanishing at the poles. Starting from the field from Eq. (4), we can double the excitation frequency along one direction, or

$$
\begin{aligned}
B_x &= B_h \cos(2\pi f t), \\
B_y &= B_h \sin(4\pi f t), \\
B_z &= B_v.
\end{aligned}
\tag{5}
$$

This is the equation of a Lissajous figure with a ratio of 1:2, as sketched in Fig. 1a. Such a field describes a figure-8 pattern in the horizontal plane, or lemniscate. The resulting trajectory is shown in Fig. 5a, over 20 periods $T = 1/f$ of the horizontal magnetic field.

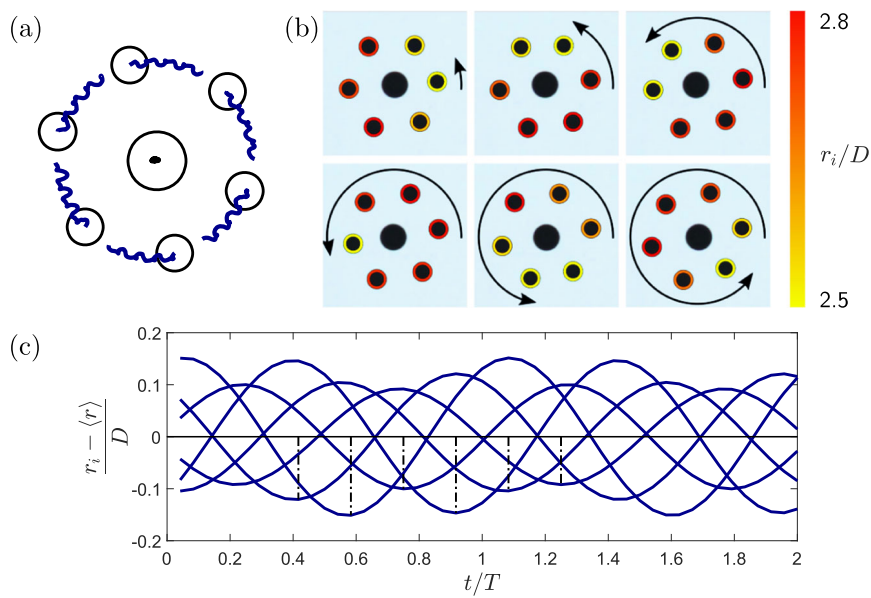

**Fig. 4 Rotation of a self-assembly using a precessing field. a** The precessing field described by Eq. (4) induces a rotation of {1,6,0}. The field has a frequency $f = 1.5$ Hz, a horizontal amplitude $B_h = 0.74$ mT and a vertical magnitude $B_z = 4.9$ mT. The trajectories are shown over four periods $T$ of the precessing magnetic field. **b** Snapshots of {1,6,0} separated by $T/6$. The peripheral particles are colored according to their distance $r_i$ to the central one, so that the closest is yellow and the farthest red. Arrows indicate the anticlockwise propagation of the deformation wave. **c** Evolution for each bead of distance $(r_i(t) - \langle r \rangle)/D$, where $D = 500$ μm is the diameter of the peripheral particles, over two periods $T$ of the magnetic field. A filter was applied at the maximum of the Fourier transform for each trajectory. Dot dashed vertical lines indicate the successive minima, evidencing a phase shift of $2\pi/6$ between neighboring beads.

A similar trajectory is shown in Supplementary Movie 2 for the larger assembly {2,8,0}. For translational motion, the typical precession frequency $f$ is taken between $10^{-1}$ Hz and 2 Hz. In Fig. 5b, two successive waves of deformation on each side of the assembly can be seen. The yellow circle, which denotes the peripheral particle closest to the central one, moves from back to front. The system is translating at a speed of ~$10^{-1}D/T$. This is comparable to the speeds that were obtained in the much simpler three-particle systems, ranging from $10^{-2}D/T$ in the collinear case to five $10^{-1}D/T$ in the triangular one[26].

Figure 5c displays the dimensionless distances $(r_i(t) - \langle r \rangle)/D$ as a function of time, for two periods $T = 1/f$. The frequency spectrum of the distances $r_i(t)$ shows peaks at $f$ and $2f$. To better visualize the movement of each bead, a filter at the frequency $f$ has been applied (see "Methods" section). Two groups of three particles can be distinguished that correspond to the left and right sides of the assembly, their trajectories shown respectively in green and in blue. Each side is successively affected by the passage of the magnetic field. The deformation wave moves from back to front, with a shift of ~$T/8$ between the oscillations of the three particles. We use the same color code in Fig. 5a, c, where the beads from the back to the front of the assembly are colored light to dark, allowing to see the propagation of the wave.

The waves shown in Fig. 5 cause a translation toward the positive $y$-axis. To cause the assembly to swim toward the negative $y$-axis, the motion of the field must proceed in the opposite direction, for instance by inverting the sign of $B_y$ in Eq. (5). It is possible to swim in any direction by changing the orientation of the lemniscate in the plane of the interface. For instance, inverting the $x$ and $y$ components in Eq. (5) causes a translation along the $x$-axis. By construction, any trajectory in the plane is therefore possible, following the same remote control strategy as in previously studied, low-particle-count cases[24].

Since the particles are mostly immersed in water, i.e., 80% of their volume is immersed, we will consider their Reynolds number $\mathrm{Re} = \rho v D/\eta$, with $\rho$ and $\eta$ the density and viscosity of water. For all motions discussed above, Re is between $10^{-2}$ and $10^{-1}$. To eliminate any possible influence of inertia, additional experiments were performed on a water–glycerol mixture, increasing the viscosity by a factor 10 ($10 \pm 0.5$ mPa). This allows to reach Reynolds numbers down to $10^{-3}$. The behavior of the assembly is generally unchanged, as seen in Supplementary Movie 3. The only noticeable difference is in the swimming speed, which is ~0.8 times the speed on water. Such a small effect of a tenfold increase in viscous drag might seem surprising; however, this is not uncommon in microswimmers, as the swimming stroke can also be affected by the change in viscosity[50]. Comparatively, few-particle magnetocapillary swimmers were considerably more affected by a change in viscosity[15].

**Locomotion mechanism.** As with ciliates, both individual and collective effects between the peripheral particles could lead to locomotion. First, consider a single peripheral particle $i$ and its position relative to the central sphere $\mathbf{r}_i$, as defined in Fig. 1a, that we will express in terms of polar coordinates $r_i$ and $\alpha_i$. From Eq. (2), we see that under a horizontal external field $\mathbf{B_x}$, the minimum of energy corresponds to the situation, where $\mathbf{r}_i$ and $\mathbf{B_x}$ are parallel. Because of the presence of other particles in the assembly, particle $i$ can only move around its equilibrium position. Therefore, $\alpha_i$ will oscillate, following the direction of the horizontal field, when $\mathbf{r}_i$ and $\mathbf{B_x}$ are close to being aligned. Concerning distance $r_i$, as discussed in Fig. 3, it becomes smaller when $\mathbf{B_x}$ and $\mathbf{r}_i$ are parallel. The combination of these two effects is a trajectory that resembles the effective and recovery strokes of cilia, as shown in Fig. 6a. The exact shape of the cycle in $(r, \alpha)$ can vary, as the interactions are nonlinear and the neighboring particles also have an influence. However the basic characteristics of an effective stroke far from the central sphere ($\mathbf{B_x}$ and $\mathbf{r}_i$ anti-parallel) and a recovery stroke closer to it ($\mathbf{B_x}$ and $\mathbf{r}_i$ parallel) are consistently seen. Because the recovery stroke always follows the direction of the wave, so does the induced motion. In that case, the direction of the wave is said to be antiplectic. By contrast, *Volvox* typically generates symplectic waves[47].

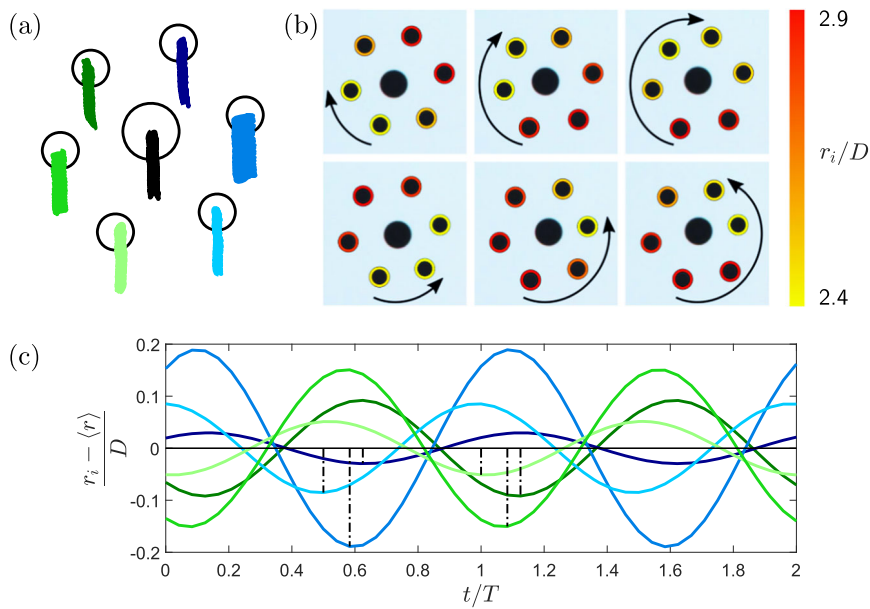

**Fig. 5 Translation of a self-assembly using a Lissajous figure. a** When the horizontal magnetic field follows the figure-8 pattern described by Eq. (5), a translation of {1,6,0} is induced. The field has a frequency $f = 1.5$ Hz (which is doubled along the $y$-direction), a horizontal amplitude $B_h = 0.74$ mT and a vertical magnitude $B_z = 4.9$ mT. The trajectories are shown over 20 periods $T = 1/f$ of the magnetic field. The trajectories are not perfectly symmetric due to slight differences in the particles' properties. **b** Snapshots of {1,6,0} separated by $T/8$. Note that $3T/8$ and $7T/8$ are not shown as they correspond to the zero of the horizontal field. The peripheral particles are colored according to their distance $r_i$ to the central one, so that the closest is yellow and the farthest red. Arrows show the propagation of the two deformation waves, left and right. **c** Evolution for each bead of distance $\langle r_i(t) - \langle r \rangle \rangle / D$, where $D = 500$ μm is the diameter of the peripheral particles, over two periods $T$ of the horizontal magnetic field, filtered at the maximum of the Fourier transform. The color code corresponds to the color in **a** to emphasize the direction of the two waves.

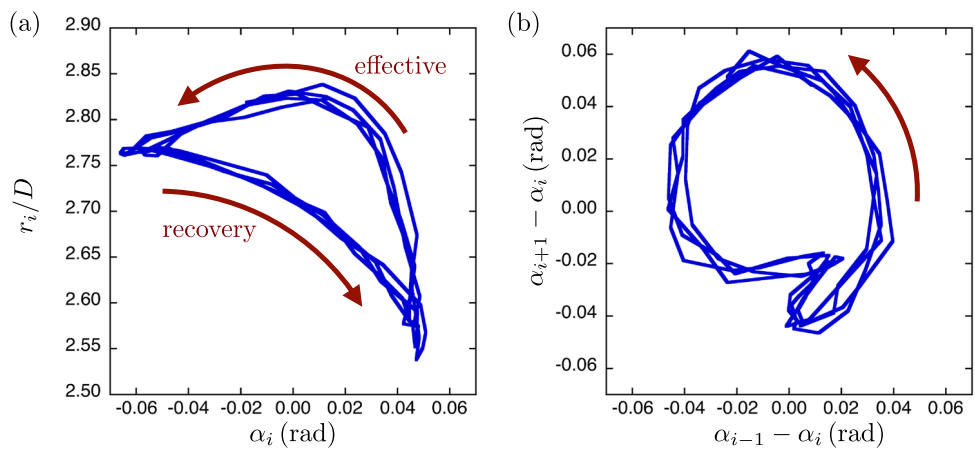

**Fig. 6 Nonreciprocal motion of the peripheral particles.** Nonreciprocal cycles in {1,6,0}, which is subjected to a precessing field with a frequency $f = 1.5$ Hz, a horizontal amplitude $B_h = 0.74$ mT and a vertical magnitude $B_z = 4.9$ mT. Trajectories are shown for four periods $T$ of the horizontal field. **a** Position of a single peripheral particle in polar coordinates, where the distance from the central bead $r_i$ is rescaled by the diameter of the bead $D$, and the angle $\alpha_i$ is taken from an arbitrary position. A single peripheral particle experiences an effective stroke, then a recovery stroke closer to the central sphere. **b** Neighboring particles move out of phase, shown here by measuring the angle $\alpha_i$ of a particle relative to its two first neighbors. The trajectory is close to a circle, which for harmonic oscillations would correspond to a phase shift of $\pi/2$.

To visualize how a phase shift between neighboring particles can also lead to locomotion, Fig. 6b compares the angle $\alpha_i$ of a particle with the angles of its neighbors. One could similarly measure the distance between neighboring particles; however, the angle is slightly easier to interpret as it is decoupled from the variation of $r_i$. When three spheres on a line oscillate, they produce a net flow proportional to the sine of the phase difference $\Delta\phi$ between the oscillations of each neighboring pair, as well as the product of their amplitudes[51,52]. In the plane defined by their interdistances, this corresponds to the area of the enclosed cycle. The direction of the induced motion of the assembly is determined by the sign of $\sin \Delta\phi$ (ref. [51]), which in our case is imposed by the direction of the metachronal wave. As this phase difference, typically given by $2\pi/N_t$, where $N_t$ is the number of peripheral particles, is always smaller than $\pi$, the induced motion follows the direction of the wave.

In the case of an assembly, such as {1,6,0}, the peripheral particles are, of course, not aligned. The angle between two neighboring pairs is, on average, 60 degrees. The effect that this angle has on a collinear three-particle swimmer is to change the orientation of the velocity, causing a rotation whose radius depends on the deformation amplitude and the angle, with a local maximum of angular displacement close to 60 degrees[53]. The effect of this angular displacement likely depends on the symmetry considered. In the translation case, the contributions from each side of the assembly of this angular displacement will cancel out over one period. However, the strokes left and right of the assembly are in phase opposition. This might explain the slight periodic rocking motion of the central particle that can be seen in Fig. 5a, as the rotation radius likely does not match the radius of the assembly. Concerning the rotation regime, both the tangential speed and this angular displacement would cause a rotation in the same direction[53]. Whether this situation is favorable compared to a purely tangential contribution remains an open question.

One might wonder whether the individual stroke and the phase shift between neighboring particles contribute equally to locomotion, or if one effect dominates over the other. As discussed above, assuming a simplified geometry, the net fluid flow is in both situations proportional to the area enclosed by the cycles shown in Fig. 6. When expressed in units of angles and length ratios, the areas of these two types of cycles are comparable. However, the prefactor that links the nonreciprocal cycles with the fluid flow is typically given by the geometry of the system[45,54]. To quantitatively compare the net quantity of fluid displaced in each case would require a comprehensive theoretical study taking fully into account the geometry of the system. Furthermore, in both situations, the motion always follows the direction of the metachronal wave. On the one hand, the recovery stroke corresponds to the moment where $r_i$ is the smallest, which happens when $\mathbf{B_x}$ and $\mathbf{r}_i$ are parallel. On the other, the phase difference between neighboring particles, which determines the swimming direction in a three-particle swimmer, is also imposed by the course of the wave and would lead to a motion in the same direction. Therefore, while the motion observed likely is, as in ciliates, a result of both the individual strokes and the phase difference between neighbors, identifying their relative contribution would require a quantitative theoretical study, as the effects cannot be separated experimentally. However, looking at assemblies of different sizes might give additional insights into the mechanism, as we will see in the next section.

**Universality of the approach**. While the assembly {1,6,0} has been used as an example throughout this paper, the wave-based approach has the advantage of remaining general. For instance, Supplementary Movie 2 shows the translational motion of the asymmetric raft {2,8,0} from Fig. 1f. In Fig. 7, we measure the speed $v$ of seven different structures, which depends on the total number of particles $N$. Structures from $N = 6$ to $N = 19$ are shown, corresponding respectively to {1,5,0} and {1,6,12}. These structures all have a core of one or several larger particles, and one or two layers of smaller particles at their periphery. A much wider range of configurations has been successfully tested, including assemblies where $N < 6$, monodisperse assemblies, and asymmetric ones. However, to study the relation between $N$ and $v$, we will only consider the most common and stable configurations. Particles in a magnetocapillary assembly tend to pack in a hexagonal or pentagonal symmetry[19]. While a purely hexagonal lattice would be expected on a flat surface, the particles cause a local curvature of the interface. Other assemblies include ones with incomplete layers, and rarer configurations such as the octagonal {1,8,0} in Fig. 1c. These structures are often metastable, causing the particles to reorganize under variations of the applied field. As abrupt configuration changes and crowding effects can affect the average speed, only the more stable assemblies where the central particles are surrounded by five or six neighbors were considered in Fig. 7.

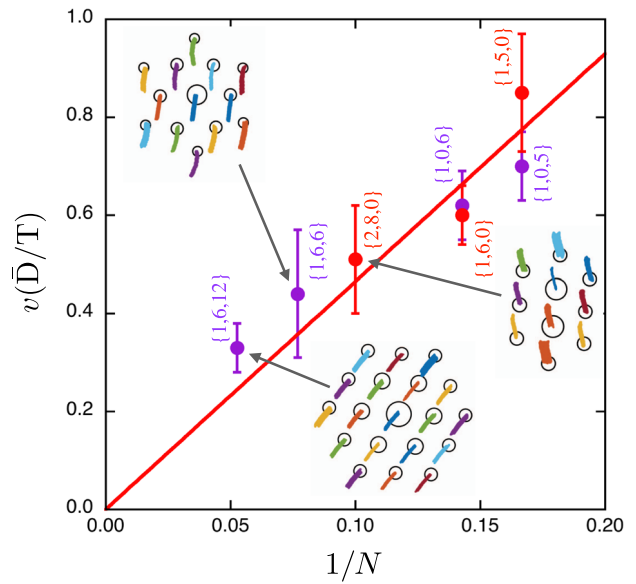

**Fig. 7 Speed of assemblies of different size.** Average speed $v$ (in units of the average diameter over all beads in the assembly $\bar{D}$ per period of oscillation $T$) as a function of $1/N$ for assemblies of various sizes. The amplitude of the horizontal field was kept at $B_h = 0.5$ mT, while the vertical field $B_z$ was adjusted between 4.5 and 6.3 mT to keep the distance between particles constant across self-assemblies. For each data point, six experiments have been done where the frequency of the horizontal field $f$ was varied between 0.5 and 2 Hz. When rescaled by $T$, we observed that $v$ is independent of $f$ for this range of frequencies. The average speed is plotted, and error bars correspond to the standard deviation. The line emphasizes the $1/N$ scaling expected for large systems. Examples of particle trajectories are also shown as insets.

The average speed $v$ is given in diameter $\bar{D}$ per period of oscillation $T$, where $\bar{D} = \sum N_i D_i / N$ corresponds to the average diameter of the beads. The experimental parameters were identical for the various assemblies, with two exceptions. First, the vertical field $B_z$ was adjusted to keep the distance between particles constant, as large assemblies tend to be more compact. Secondly, to minimize the influence of resonances that can appear in magnetocapillary interactions[25], every data point in Fig. 7 was averaged over several values of $f$ between 0.5 and 2 Hz. When $v$ is expressed in diameters per period, it is roughly independent of frequency $f$ for the frequencies considered herein. Insets show typical assemblies and their trajectories. The red and purple dots distinguish assemblies where $N_3 = 0$ and $N_3 \neq 0$, respectively. For instance, the two dots on the top right corner of Fig. 7 represent {1,0,5} for the purple one, and {1,5,0} for the red one. The reason for changing the size of the peripheral particles is to change the magnitude of some forces in the system, most notably the hydrodynamic and viscous forces. However, we can see that the points overlap for the same $N$, if we rescale the speed by the average diameter of the assembly. The speed $v$ grows linearly with $1/N$, meaning that the largest rafts tend to be less efficient under fixed conditions.

Though describing analytically, the dynamics of large assemblies would be complex due to the many degrees of freedom, it is possible to justify the $1/N$ scaling from Fig. 7. The thrust provided by the cooperative motion of peripheral particles can be written $F_t \sim N_t e$, where $N_t < N$ is the number of these particles acting as motors (or cilia) and $e$ is the efficiency of each motor. To simplify the discussion, we will consider that $N_t$ is given by the number of beads in the outermost layer of the assembly. This is equivalent to considering that neighboring particles on successive layers act as a single cilium, as their motion is coupled. Moreover, we will make the assumption that, of the two mechanisms described in the previous section, the effect of the phase shift between neighboring particles is dominant. In this case, we expect the efficiency $e$ of each motor to be proportional to the area of the nonreciprocal cycles as presented in Fig. 6b, and therefore to the sine of the phase shift between adjacent beads. For small phase differences, we find $e \sim 1/N_t$. In the particular case of precessing fields (Eq. (4)), the phase shift is simply given by $2\pi/N_t$. As a result, the total thrust $F_t$ is independent on the total number of particles $N$, and of the number of motors $N_t$. At low Reynolds number, the total thrust $F_t$ should be counterbalanced by the viscous forces $F_\eta$ acting on all particles of the structure, i.e., $F_t = F_\eta$ with $F_\eta \sim N\nu$. As a result, the average translation speed of the assembly is given by

$$v(N) = v_0/N \ . \tag{6}$$

This scaling is in agreement with the data from Fig. 7, and a fit provides the value $v_0 = 4.6 \pm 0.2$ in average diameter $\bar{D}$ per period $T$. Had we made the assumption that the stroke of a single peripheral particle was the dominant mechanism, efficiency $e$ would have been independent on $N_t$. This would have led to the scaling $v = v_0' N_t / N$. Indeed, even if crowding could affect the dynamics of a particle, the number of first neighbors is relatively constant for the motor particles in the structures considered here, i.e., with a complete outermost layer of particles on a hexagonal or pentagonal lattice. The results from Fig. 7 therefore suggest that the effect of the phase shift is dominant for the assemblies considered here.

The slope $v_0$ is likely dependent on the experimental parameters, including particle-to-particle distance and amplitude $B_h$. Future studies could aim at maximizing $v_0$ by changing the properties of the horizontal field. In Eq. (5), one could vary the relative amplitude as well as the phase difference between $B_x$ and $B_y$, change the frequency ratio, or change the waveform altogether. For instance, a quadrifolium curve might generate two successive waves on each side of the assembly. One could also create a piecewise function such that the horizontal field never goes through zero. Indeed, when the field is interrupted, the assembly travels by a dimensionless coasting distance given by $d/D \sim \mathrm{Re}\ \rho_s/\rho \simeq 10^{-1}$, where $\rho_s$ is the density of the particles[12]. This means that any interruption in the waves is essentially lost, as coasting is very small at low Reynolds number.

## Discussion

Metachronal waves are abundantly found in nature, from unicellular organisms to insects and crustaceans. They provide a way to coordinate a large number of appendages in order to generate locomotion efficiently. Using self-organized rafts of magnetic particles, we demonstrated how to produce rotational and translational motion at low Reynolds number through metachronal waves in an artificial system. Precessing magnetic fields induce waves of deformations that propagate in the system in a controlled way. These local deformations are triggered by a uniform magnetic field, owing to the properties of the dipole–dipole interaction. By contrast with most magnetically powered swimmers, this strategy remains general and works on assemblies of various sizes and symmetries. As the system is self-assembled, no microfabrication is needed. However, this approach could possibly also be applied to magnetic artificial cilia attached to a body. Indeed, similarly to the particles used herein, the response of magnetic cilia typically depends on the relative orientation with the field[38–40]. The strategy of propagating deformation waves using Lissajous curves might therefore be applicable.

This swimming strategy is not only general, but quite efficient, approaching speeds of about one particle diameter per period of oscillation. However, optimizing the swimming speed was not the primary objective of this paper. Further research could focus on engineering faster swimmers by changing the parameters of the magnetic field, as well as the properties of the structures. One could perform experiments with particles of different shapes, sizes, and wetting properties. Different types of liquid interfaces can also be explored, such as oil–water interfaces or soap films. The conceptual simplicity of the approach presented in this paper leaves many possibilities open, from more complex self-assemblies to systems with microfabricated cilia, from the most fundamental aspects of biomimetic propulsion to versatile microsystems engineered to perform various tasks at the milli and microscale.

## Methods

The experimental setup is the following. A glass container is filled with either water or, to change the viscosity, a glycerol and water mixture stirred for 48 h, and its viscosity measured with a rheometer. It is placed in the center of a large tri-axis Helmholtz system of coils. By injecting current into the coils, spatially uniform magnetic induction fields can be generated in any direction inside the system. The $z$ coil is used to generate a constant, vertical magnetic field $\mathbf{B}_z$ and is therefore fed by a DC current generator. The $x$ and $y$ coils are connected to amplifiers and a multichannel arbitrary function generator (AFC), which allows to generate oscillating fields in the plane of the interface. The AFC is commanded by a computer for better control of the parameters, allowing for instance to compensate the Earth's magnetic field and to change the oscillation direction quickly. The magnitude of $\mathbf{B}_z$ hardly exceeds 5 mT, while the horizontal amplitudes are usually smaller by a factor 2 to prevent contact between the particles. Said particles are precision spheres made of either martensitic stainless steel (AISI 420) or low alloy martensitic chrome steel (AISI 52100). Under a magnetic field, these particles behave like an ideal soft magnet, with very little remanence and coercivity[25]. The exact diameters used are 397, 500, and 793 μm. The spheres are highly spherical and ~7.8 times denser than water.

The container is made of glass and covered with a glass lid to avoid contamination. The glass is coated with a transparent conducting oxide and connected to earth, in order to prevent the build up of electric charge. Prior to each experiment, the tank and lid are carefully washed, and the particles and water replaced. Consistency is essential when filling the tank and placing the particles to avoid variations in the meniscus of the bath and the contact line of the particles. The bath is lit from below and filmed from the top, using a camera equipped with a macro lens. To measure the position of the particles, they are tracked through a circle Hough transform. In Figs. 4 and 5, the trajectories were filtered to make the comparison between different particles easier. Indeed, the trajectories can be quite nonlinear and there can be slight differences between particles due to variations in their magnetic properties and wetting. To single out the dominant part of each signal, we performed a Fourier transform and multiplied each spectrum by a filter equal to 1 at its maximum and 0 elsewhere. This allows to identify the phase difference between the various trajectories more clearly.

## Data availability
The data that support the findings of this study are available from the corresponding author upon reasonable request.

## Code availability
The code used in this study is available from the corresponding author upon reasonable request.

Received: 6 January 2020; Accepted: 29 May 2020;
Published online: 19 June 2020

## References
1. Zhang, L. et al. Artificial bacterial flagella: fabrication and magnetic control. *Appl. Phys. Lett.* **94**, 064107 (2009).
2. Williams, B. J., Anand, S. V., Rajagopalan, J. & Saif, M. T. A. A self-propelled biohybrid swimmer at low reynolds number. *Nat. Commun.* **5**, 3081 (2014).
3. Gao, W. et al. Bioinspired helical microswimmers based on vascular plants. *Nano Lett.* **14**, 305–310 (2013).

4. Huang, H.-W. et al. Adaptive locomotion of artificial microswimmers. *Sci. Adv.* **5**, eaau1532 (2019).

5. Lancia, F. et al. Reorientation behavior in the helical motility of light-responsive spiral droplets. *Nat. Commun.* **10**, 1–8 (2019).

6. Dai, B. et al. Programmable artificial phototactic microswimmer. *Nat. Nanotechnol.* **11**, 1087 (2016).

7. Popescu, M. N., Uspal, W. E. & Dietrich, S. Self-diffusiophoresis of chemically active colloids. *Eur. Phys. J. Spec. Top.* **225**, 2189–2206 (2016).

8. Ahmed, D. et al. Selectively manipulable acoustic-powered microswimmers. *Sci. Rep.* **5**, 9744 (2015).

9. Rao, K. J. et al. A force to be reckoned with: a review of synthetic microswimmers powered by ultrasound. *Small* **11**, 2836–2846 (2015).

10. Lozano, C., TenHagen, B., Löwen, H. & Bechinger, C. Phototaxis of synthetic microswimmers in optical landscapes. *Nat. Commun.* **7**, 12828 (2016).

11. Palagi, S. et al. Structured light enables biomimetic swimming and versatile locomotion of photoresponsive soft microrobots. *Nat. Mat.* **15**, 647 (2016).

12. Lauga, E. & Powers, T. R. The hydrodynamics of swimming microorganisms. *Rep. Prog. Phys.* **72**, 096601 (2009).

13. Peyer, K. E., Zhang, L. & Nelson, B. J. Bio-inspired magnetic swimming microrobots for biomedical applications. *Nanoscale* **5**, 1259–1272 (2013).

14. Snezhko, A. & Aranson, I. S. Magnetic manipulation of self-assembled colloidal asters. *Nat. Mater.* **10**, 698–703 (2011).

15. Lumay, G., Obara, N., Weyer, F. & Vandewalle, N. Self-assembled magnetocapillary swimmers. *Soft Matter* **9**, 2420–2425 (2013).

16. Martinez-Pedrero, F. & Tierno, P. Magnetic propulsion of self-assembled colloidal carpets: efficient cargo transport via a conveyor-belt effect. *Phys. Rev. Appl.* **3**, 051003 (2015).

17. Golosovsky, M., Saado, Y. & Davidov, D. Self-assembly of floating magnetic particles into ordered structures: a promising route for the fabrication of tunable photonic band gap materials. *Appl. Phys. Lett.* **75**, 4168–4170 (1999).

18. Wen, W., Zhang, L. & Sheng, P. Planar magnetic colloidal crystals. *Phys. Rev. Lett.* **85**, 5464–5467 (2000).

19. Vandewalle, N. et al. Symmetry breaking in a few-body system with magnetocapillary interactions. *Phys. Rev. E* **85**, 041402 (2012).

20. Vandewalle, N., Obara, N. & Lumay, G. Mesoscale structures from magnetocapillary self-assembly. *Eur. Phys. J. E* **36**, 1–6 (2013).

21. Grzybowski, B., Stone, H. A. & Whitesides, G. M. Dynamic self-assembly of magnetized, millimetre-sized objects rotating at a liquid-air interface. *Nature* **405**, 1033–1036 (2000).

22. Wang, Q., Yang, L., Wang, B., Yu, E., Yu, J. & Zhang, L. Collective behavior of reconfigurable magnetic droplets via dynamic self-assembly. *ACS Appl. Mater. Interfaces* **11**, 1630–1637 (2018).

23. Chinomona, R., Lajeunesse, J., Mitchell, W. H., Yao, Y. & Spagnolie, S. E. Stability and dynamics of magnetocapillary interactions. *Soft Matter* **11**, 1828–1838 (2015).

24. Grosjean, G. Remote control of self-assembled microswimmers. *Sci. Rep.* **5**, 16035 (2015).

25. Lagubeau, G. et al. Statics and dynamics of magnetocapillary bonds. *Phys. Rev. E* **93**, 053117 (2016).

26. Grosjean, G., Hubert, M. & Vandewalle, N. Magnetocapillary self-assemblies: locomotion and micromanipulation along a liquid interface. *Adv. Colloid Interface Sci.* **255**, 84–93 (2018).

27. Grosjean, G., Hubert, M., Collard, Y., Pillitteri, S. & Vandewalle, N. Surface swimmers, harnessing the interface to self-propel. *Eur. Phys. J. E* **41**, 137 (2018).

28. Sukhov, A. et al. Optimal motion of triangular magnetocapillary swimmers. *J. Chem. Phys.* **151**, 124707 (2019).

29. Grosjean, G. et al. Capillary assemblies in a rotating magnetic field. *Soft Matter* **15**, 9093–9103 (2019).

30. Osterman, N. & Vilfan, A. Finding the ciliary beating pattern with optimal efficiency. *Proc. Natl Acad. Sci. USA* **108**, 15727–15732 (2011).

31. Elgeti, J. & Gompper, G. Emergence of metachronal waves in cilia arrays. *Proc. Natl Acad. Sci. USA* **110**, 4470–4475 (2013).

32. Alben, S., Spears, K., Garth, S., Murphy, D. & Yen, J. Coordination of multiple appendages in drag-based swimming. *J. R. Soc. Interface* **7**, 1545–1557 (2010).

33. Osborn, K. J., Haddock, S. H., Pleijel, F., Madin, L. P. & Rouse, G. W. Deep-sea, swimming worms with luminescent bombs. *Science* **325**, 964–964 (2009).

34. Wilson, D. M. Insect walking. *Annu. Rev. Entomol.* **11**, 103–122 (1966).

35. Lauga, E. & Bartolo, D. No many-scallop theorem: collective locomotion of reciprocal swimmers. *Phys. Rev. E* **78**, 030901 (2008).

36. Childress, S. *Mechanics of Swimming and Flying*, Vol. 2 (Cambridge University Press, 1981).

37. Narematsu, N., Quek, R., Chiam, K. H. & Iwadate, Y. Ciliary metachronal wave propagation on the compliant surface of Paramecium cells. *Cytoskeleton* **72**, 633–646 (2015).

38. Vilfan, M. et al. Self-assembled artificial cilia. *Proc. Natl Acad. Sci. USA* **107**, 1844–1847 (2010).

39. Meng, F., Matsunaga, D., Yeomans, J. & Golestanian, R. Magnetically-actuated artificial cilium: a simple theoretical model. *Soft Matter* **15**, 3864–3871 (2019).

40. Poty, M., Weyer, F., Grosjean, G., Lumay, G. & Vandewalle, N. Magnetoelastic instability in soft thin films. *Eur. Phys. J. E* **40**, 29 (2017).

41. Haynes, G. C., Rizzi, A. A. Gaits and gait transitions for legged robots. In *Proceedings 2006 IEEE International Conference on Robotics and Automation, 2006. ICRA 2006,* 1117–1122 (IEEE, 2006).

42. Palagi, S., Jager, E. W., Mazzolai, B. & Beccai, L. Propulsion of swimming microrobots inspired by metachronal waves in ciliates: from biology to material specifications. *Bioinspiration Biomim.* **8**, 046004 (2013).

43. Kralchevsky, P. A. & Nagayama, K. Capillary forces between colloidal particles. *Langmuir* **10**, 23–36 (1994).

44. Vella, D. & Mahadevan, L. The Cheerios effect. *Am. J. Phys.* **73**, 817–825 (2005).

45. Vilfan, A. & Jülicher, F. Hydrodynamic flow patterns and synchronization of beating cilia. *Phys. Rev. Lett.* **96**, 058102 (2006).

46. Vilfan, A. Generic flow profiles induced by a beating cilium. *Eur. Phys. J. E* **35**, 72 (2012).

47. Brumley, D. R., Polin, M., Pedley, T. J. & Goldstein, R. E. Hydrodynamic synchronization and metachronal waves on the surface of the colonial alga Volvox carteri. *Phys. Rev. Lett.* **109**, 268102 (2012).

48. Pedley, T. J., Brumley, D. R. & Goldstein, R. E. Squirmers with swirl: a model for volvox swimming. *J. Fluid Mech.* **798**, 165–186 (2016).

49. Ueki, N., Matsunaga, S., Inouye, I. & Hallmann, A. How 5000 independent rowers coordinate their strokes in order to row into the sunlight: phototaxis in the multicellular green alga Volvox. *BMC Biol.* **8**, 103 (2010).

50. Jayant, P., Merchant, L., Krüger, T., Harting, J. & Smith, A. S. Setting the pace of microswimmers: when increasing viscosity speeds up self-propulsion. *N. J. Phys.* **19**, 053024 (2017).

51. Najafi, A. & Golestanian, R. Simple swimmer at low Reynolds number: three linked spheres. *Phys. Rev. E* **69**, 062901 (2004).

52. Grosjean, G., Hubert, M., Lagubeau, G. & Vandewalle, N. Realization of the Najafi-Golestanian microswimmer. *Phys. Rev. E* **94**, 021101 (2016).

53. Ledesma-Aguilar, R., Löwen, H. & Yeomans, J. M. A circle swimmer at low Reynolds number. *Eur. Phys. J. E* **35**, 70 (2012).

54. Golestanian, R. & Ajdari, A. Analytic results for the three-sphere swimmer at low Reynolds number. *Phys. Rev. E* **77**, 036308 (2008).

## Acknowledgements
This work was financially supported by the FNRS (grant PDR T.0129.18). G.G. thanks FRIA for financial support. This project has received funding from the European Union's Horizon 2020 research and innovation programme under the Marie Skłodowska-Curie Grant Agreement No. 754411.

## Author contributions
The experiments were performed by Y.C. G.G. and N.V. provided the interpretations of the results. Y.C., G.G., and N.V. all contributed to redaction of the manuscript.

## Competing interests
The authors declare no competing interests.

## Additional information
**Supplementary information** is available for this paper at https://doi.org/10.1038/s42005-020-0380-9.

**Correspondence** and requests for materials should be addressed to Y.C.

**Reprints and permission information** is available at http://www.nature.com/reprints

**Publisher's note** Springer Nature remains neutral with regard to jurisdictional claims in published maps and institutional affiliations.